\documentclass[preprint,aps]{revtex4}

\begin{document}

\title{Comment on \textquotedblleft Exact results for survival probability
in the multistate Landau-Zener model\textquotedblright}
\author{B. E. Dobrescu}
\affiliation{Department of Physics, Texas A\&M University, College Station, TX
77843-4242, USA }
\author{N. A. Sinitsyn}
\affiliation{Department of Physics, University of Texas at Austin, Austin TX 78712-1081,
USA}

\begin{abstract}
We correct the proof of Brundobler-Elser formula (BEF) provided in [2004 
\textit{J. Phys. B: At. Mol. Opt. Phys.} \textbf{37} 4069] and continued in
Appendix of [2005 \textit{J. Phys. B: At. Mol. Opt. Phys.} \textbf{38} 907].
After showing that some changes of variables employed in these articles are
used erroneously, we propose an alternative change of variables which solves
the problem. In our proof, we reveal the connection between the BEF for a
general $N$-level Landau-Zener system and the exactly solvable bow-tie
model. The special importance of the diabatic levels with maximum/minimum
slope is emphasized throughout.
\end{abstract}

\maketitle

\section{Introduction}

In Ref. \cite{VO1} Volkov and Ostrovsky (referred to below as V-O) proposed
a rigorous mathematical proof of the Brundobler-Elser formula (BEF) \cite{BE}
for a general $N$-level Landau-Zener (LZ) system. The approach of \cite{VO1}
was to consider a formal solution in terms of an infinite series in powers
of the coupling constants, with further calculation of every term and
summation of the resulting algebraic series. Following this strategy, these
authors have demonstrated the exact cancellations of some quantum
paths and have reduced the problem to the calculation of a sub-series $(5.1)$
(the numbering of formulas is as in \cite{VO1}).

In the calculation of this remaining sub-series $(5.1)$ in \cite{VO1} V-O
made the error of assuming a wrong domain of integration for their variables 
$x_{j}$ (see Eq. $\left( 5.2\right) $ of \cite{VO1}).

In Ref. \cite{VO2}, whose main text is dedicated to a different subject, V-O
included an Appendix in which a new attempt is made at calculating the
sub-series $(5.1)$. After acknowledging the error made in \cite{VO1}, V-O
proceeded to employ a different change of variables, previously proposed by
Kayanuma and Fukuchi in \cite{KayaF}.

We would like to point out that this derivation contained in the Appendix of 
\cite{VO2} is still erroneous. After Eq. (A.4) (in the notation of Ref. \cite%
{VO2}) the authors claim that the integrand is symmetrical under the
permutation of variables $x_{0},\ldots ,x_{m-1}$. It is, however,
straightforward to check, by direct examination of the terms containing the
4-th power in the coupling constants $V_{i}$, that this is not true; nor is
it symmetrical under exchange of pairs of variables $(x_{0},y_{0}),\ldots
,(x_{m-1},y_{m-1})$.

The following example will support these considerations: the term
corresponding to $l=m=2$ in Eq. $\left( 5.1\right) $ of \cite{VO1} reads 
\begin{eqnarray}
&&\sum_{k_{1}\neq 1}^{N}\left\vert V_{1k_{1}}\right\vert ^{2}\sum_{k_{2}\neq
1}^{N}\left\vert V_{1k_{2}}\right\vert ^{2}\int_{-\infty }^{\infty
}dt_{4}\int_{-\infty }^{t_{4}}dt_{3}\int_{-\infty
}^{t_{3}}dt_{2}\int_{-\infty }^{t_{2}}dt_{1}  \label{art1-1} \\
&&\times \exp \left[ i\left( \varepsilon _{1}-\varepsilon _{k_{1}}\right)
\left( t_{4}-t_{3}\right) +i\left( \varepsilon _{1}-\varepsilon
_{k_{2}}\right) \left( t_{2}-t_{1}\right) \right]  \nonumber \\
&&\times \exp \left[ \frac{i}{2}\left( \beta _{1}-\beta _{k_{1}}\right)
\left( t_{4}^{2}-t_{3}^{2}\right) +\frac{i}{2}\left( \beta _{1}-\beta
_{k_{2}}\right) \left( t_{2}^{2}-t_{1}^{2}\right) \right] .  \nonumber
\end{eqnarray}

Let us consider the simplest model beyond the two-level LZ system, namely
the one with $N=3$. Then, the sum above has 4 terms 
\begin{eqnarray}
I_{22} &=&\left\vert V_{12}\right\vert ^{4}\int_{-\infty }^{\infty
}dt_{4}\int_{-\infty }^{t_{4}}dt_{3}\int_{-\infty
}^{t_{3}}dt_{2}\int_{-\infty }^{t_{2}}dt_{1}  \label{art1-2} \\
&&\times \exp \left[ i\left( \varepsilon _{1}-\varepsilon _{2}\right) \left(
t_{4}-t_{3}+t_{2}-t_{1}\right) +\frac{i}{2}\left( \beta _{1}-\beta
_{2}\right) \left( t_{4}^{2}-t_{3}^{2}+t_{2}^{2}-t_{1}^{2}\right) \right] , 
\nonumber
\end{eqnarray}%
\begin{eqnarray}
I_{33} &=&\left\vert V_{13}\right\vert ^{4}\int_{-\infty }^{\infty
}dt_{4}\int_{-\infty }^{t_{4}}dt_{3}\int_{-\infty
}^{t_{3}}dt_{2}\int_{-\infty }^{t_{2}}dt_{1}  \label{art1-3} \\
&&\times \exp \left[ i\left( \varepsilon _{1}-\varepsilon _{3}\right) \left(
t_{4}-t_{3}+t_{2}-t_{1}\right) +\frac{i}{2}\left( \beta _{1}-\beta
_{3}\right) \left( t_{4}^{2}-t_{3}^{2}+t_{2}^{2}-t_{1}^{2}\right) \right] , 
\nonumber
\end{eqnarray}%
\begin{eqnarray}
I_{23} &=&\left\vert V_{12}\right\vert ^{2}\left\vert V_{13}\right\vert
^{2}\int_{-\infty }^{\infty }dt_{4}\int_{-\infty
}^{t_{4}}dt_{3}\int_{-\infty }^{t_{3}}dt_{2}\int_{-\infty }^{t_{2}}dt_{1}
\label{art1-4} \\
&&\times \exp \left[ i\left( \varepsilon _{1}-\varepsilon _{2}\right) \left(
t_{4}-t_{3}\right) +i\left( \varepsilon _{1}-\varepsilon _{3}\right) \left(
t_{2}-t_{1}\right) \right]   \nonumber \\
&&\times \exp \left[ \frac{i}{2}\left( \beta _{1}-\beta _{2}\right) \left(
t_{4}^{2}-t_{3}^{2}\right) +\frac{i}{2}\left( \beta _{1}-\beta _{3}\right)
\left( t_{2}^{2}-t_{1}^{2}\right) \right] ,  \nonumber
\end{eqnarray}%
\begin{eqnarray}
I_{32} &=&\left\vert V_{13}\right\vert ^{2}\left\vert V_{12}\right\vert
^{2}\int_{-\infty }^{\infty }dt_{4}\int_{-\infty
}^{t_{4}}dt_{3}\int_{-\infty }^{t_{3}}dt_{2}\int_{-\infty }^{t_{2}}dt_{1}
\label{art1-5} \\
&&\times \exp \left[ i\left( \varepsilon _{1}-\varepsilon _{3}\right) \left(
t_{4}-t_{3}\right) +i\left( \varepsilon _{1}-\varepsilon _{2}\right) \left(
t_{2}-t_{1}\right) \right]   \nonumber \\
&&\times \exp \left[ \frac{i}{2}\left( \beta _{1}-\beta _{3}\right) \left(
t_{4}^{2}-t_{3}^{2}\right) +\frac{i}{2}\left( \beta _{1}-\beta _{2}\right)
\left( t_{2}^{2}-t_{1}^{2}\right) \right] .  \nonumber
\end{eqnarray}

The change of variables \cite{KayaF} used by V-O in Appendix of \cite{VO2}
amounts to 
\begin{eqnarray*}
t_{1} &=&x_{1}\text{, \ \ \ \ }x_{1}\in \left( -\infty ,\infty \right) \\
t_{2} &=&x_{1}+y_{1}\text{, \ \ \ }y_{1}\in \left[ 0,\infty \right) \\
t_{3} &=&x_{2}+y_{1}\text{, \ \ \ \ }x_{2}\in \left[ x_{1},\infty \right) \\
t_{4} &=&x_{2}+y_{1}+y_{2}\text{, \ \ \ \ }y_{2}\in \left[ 0,\infty \right)
\end{eqnarray*}

The integrals $I_{22}$ and $I_{33}$ can be easily evaluated (see \cite{KayaF}%
), e.g. 
\begin{eqnarray}
I_{22} &\equiv &\left\vert V_{12}\right\vert ^{4}\int_{-\infty }^{\infty
}dx_{2}\int_{-\infty }^{x_{2}}dx_{1}\int_{0}^{\infty }dy_{2}\int_{0}^{\infty
}dy_{1}  \label{art1-6} \\
&&\times \exp \left[ i\left( \varepsilon _{1}-\varepsilon _{2}\right) \left(
y_{1}+y_{2}\right) +\frac{i}{2}\left( \beta _{1}-\beta _{2}\right) \left(
y_{1}+y_{2}\right) ^{2}+i\left( \beta _{1}-\beta _{2}\right) \left(
y_{1}x_{1}+y_{2}x_{2}\right) \right]   \nonumber \\
&=&\left\vert V_{12}\right\vert ^{4}\frac{1}{2}\int_{-\infty }^{\infty
}dx_{2}\int_{-\infty }^{\infty }dx_{1}\int_{0}^{\infty
}dy_{2}\int_{0}^{\infty }dy_{1}\times   \nonumber \\
&&\times \exp \left[ i\left( \varepsilon _{1}-\varepsilon _{2}\right) \left(
y_{1}+y_{2}\right) +\frac{i}{2}\left( \beta _{1}-\beta _{2}\right) \left(
y_{1}+y_{2}\right) ^{2}+i\left( \beta _{1}-\beta _{2}\right) \left(
y_{1}x_{1}+y_{2}x_{2}\right) \right]   \nonumber \\
&=&\left\vert V_{12}\right\vert ^{4}\frac{1}{2}\int_{0}^{\infty
}dy_{2}\int_{0}^{\infty }dy_{1}\text{ }\frac{2\pi }{\left\vert \beta
_{1}-\beta _{2}\right\vert }\delta \left( y_{1}\right) \frac{2\pi }{%
\left\vert \beta _{1}-\beta _{2}\right\vert }\delta \left( y_{2}\right)
\times   \nonumber \\
&&\times \exp \left[ i\left( \varepsilon _{1}-\varepsilon _{2}\right) \left(
y_{1}+y_{2}\right) +\frac{i}{2}\left( \beta _{1}-\beta _{2}\right) \left(
y_{1}+y_{2}\right) ^{2}\right] =\frac{\left\vert V_{12}\right\vert ^{4}}{2}%
\left( \frac{\pi }{\beta _{1}-\beta _{2}}\right) ^{2}  \nonumber
\end{eqnarray}%
\noindent and, similarly, $I_{33}=\frac{\left\vert V_{13}\right\vert ^{4}}{2}%
\left( \frac{\pi }{\beta _{1}-\beta _{3}}\right) ^{2}$.

The sum of the remaining two integrals reads 
\begin{eqnarray}
I_{23}+I_{32} &=&\left\vert V_{12}\right\vert ^{2}\left\vert
V_{13}\right\vert ^{2}\int_{-\infty }^{\infty }dx_{1}\int_{x_{1}}^{\infty
}dx_{2}\int_{0}^{\infty }dy_{2}\int_{0}^{\infty }dy_{1}  \label{eqII} \\
&&\times \left\{ \exp \left[ i\left( \varepsilon _{1}-\varepsilon
_{2}\right) y_{2}+i\left( \varepsilon _{1}-\varepsilon _{3}\right) y_{1}%
\right] \right.  \nonumber \\
&&\times \exp \left[ \frac{i}{2}\left( \beta _{1}-\beta _{2}\right) \left(
2x_{2}y_{2}+y_{2}^{2}\right) +\frac{i}{2}\left( \beta _{1}-\beta _{3}\right)
\left( 2x_{1}y_{1}+y_{1}^{2}\right) +i\left( \beta _{1}-\beta _{2}\right)
y_{1}y_{2}\right]  \nonumber \\
&&+\exp \left[ i\left( \varepsilon _{1}-\varepsilon _{3}\right)
y_{2}+i\left( \varepsilon _{1}-\varepsilon _{2}\right) y_{1}\right] 
\nonumber \\
&&\times \left. \exp \left[ \frac{i}{2}\left( \beta _{1}-\beta _{3}\right)
\left( 2x_{2}y_{2}+y_{2}^{2}\right) +\frac{i}{2}\left( \beta _{1}-\beta
_{2}\right) \left( 2x_{1}y_{1}+y_{1}^{2}\right) +i\left( \beta _{1}-\beta
_{3}\right) y_{1}y_{2}\right] \right\} .  \nonumber
\end{eqnarray}

Contrary to the claims made by V-O after Eq. (A.4) of \cite{VO2}, the
integrand of $\left( \ref{eqII}\right) $ is clearly \textit{not symmetric}
with respect to interchanging the pairs $\left( x_{1},y_{1}\right) $ and $%
\left( x_{2},y_{2}\right) $. Indeed, the terms $\left( \beta _{1}-\beta
_{2}\right) y_{1}y_{2}\neq \left( \beta _{1}-\beta _{3}\right) y_{1}y_{2}$
belong to different terms of the sum in the integrand and are left unchanged
by the permutation. Therefore, one cannot extend the domain of integration
of variable $x_{2}$ to $\left( -\infty ,+\infty \right) $ as proposed by V-O
in Eq. (A.5) of \cite{VO2}.

\bigskip

While the mathematical errors made by V-O are presented above in detail only
for completeness, there is also a more general argument as to why the
calculation of the series $(5.1)$ in both Ref. \cite{VO1} and Ref. \cite{VO2}
is erroneous. One can notice that this series represents a formal solution
for a special case of the multi-state LZ-model, in which all states interact
with only one special level with slope $\beta _{1}$. Series $(5.1)$ is the
probability amplitude for the system to remain on this special level after
all crossings if only this level has been initially populated. This
observation does \textit{not} assume at all that the slope $\beta _{1}$ is
an extremum. Therefore, if the proof provided by V-O in \cite{VO1} and \cite%
{VO2} were correct, then their summation of the series $(5.1)$ that does 
\textit{not} use the assumption of a maximal/minimal slope for the level $1$
could be applied to such a LZ-model with an \textit{arbitrary} slope $\beta
_{1}$.

However, it is well known from numerical simulations \cite{BE} of this model
that if $\beta _{1}$ is \textit{not} an extremum, then the BEF does \textit{%
not} hold, and the transition probability for $1\rightarrow 1$ depends
essentially on the parameters $\varepsilon _{i}$. Also, the exactly solvable
case of the above example in which all $\varepsilon _{i}=0$, the bow-tie
model \cite{bow-tie}, proves that the diagonal elements of the $S$ matrix
are \textit{different} from BEF \textit{except} for the levels with \textit{%
maximal/minimal} slope.

The total ignorance of this very important property of the slope $\beta _{1}$
being an \textit{extremum} in the calculation of series $(5.1)$ by V-O has
actually been the trigger that prompted our search for their mathematical
mistakes.

\bigskip

\section{Proof of BEF}

We will split the proof of BEF in three steps. The first step has been
correctly performed in \cite{VO1} and we only briefly mention here the main
results in order to introduce the notation and as a starting point for the
next steps.

It has been shown in \cite{VO1} that \textit{if} the slope $\beta _{1}$ is 
\textit{maximal/minimal} then the transition amplitude for the transition $%
1\rightarrow 1$ has the form: 
\begin{eqnarray}
S_{11} &\equiv &\langle 1|\hat{U}\left( \infty ,-\infty \right) |1\rangle
=1+\sum_{m=1}^{\infty }\left( -1\right)
^{m}\sum_{k_{1}=2}^{N}\sum_{k_{2}=2}^{N}\cdots \sum_{k_{m}=2}^{N}\left\vert
V_{1k_{1}}\right\vert ^{2}\left\vert V_{1k_{2}}\right\vert ^{2}\cdots
\left\vert V_{1k_{m}}\right\vert ^{2}\times  \nonumber \\
&&\times \int_{-\infty }^{\infty }d\tau _{1}\int_{-\infty }^{\tau _{1}}d\tau
_{2}\cdots \int_{-\infty }^{\tau _{2m-1}}d\tau _{2m}  \nonumber \\
&&\times \exp \left[ i\frac{B\left( k_{1}\right) }{2}\left( \tau
_{1}^{2}-\tau _{2}^{2}\right) +i\frac{B\left( k_{2}\right) }{2}\left( \tau
_{3}^{2}-\tau _{4}^{2}\right) +\cdots +i\frac{B\left( k_{m}\right) }{2}%
\left( \tau _{2m-1}^{2}-\tau _{2m}^{2}\right) \right]  \nonumber \\
&&\times \exp \left[ iE\left( k_{1}\right) \left( \tau _{1}-\tau _{2}\right)
+iE\left( k_{2}\right) \left( \tau _{3}-\tau _{4}\right) +\cdots +iE\left(
k_{m}\right) \left( \tau _{2m-1}-\tau _{2m}\right) \right] ,  \label{fiveone}
\end{eqnarray}%
where 
\begin{equation}
B\left( k_{j}\right) \equiv \beta _{1}-\beta _{k_{j}}\text{ \ \ and \ \ }%
E\left( k_{j}\right) \equiv \varepsilon _{1}-\varepsilon _{k_{j}}\text{, \ \
\ \ \ }j=1,2,\ldots ,m\text{.}  \label{eq2}
\end{equation}%
$\hat{U}$ is the time-evolution operator and $N$ is the number of levels in
the system.

Series (\ref{fiveone}) is equivalent to the series $(5.1)$ in \cite{VO1}. 
From this point on, Volkov and Ostrovsky do \textit{not} use the property of 
$\beta _{1}$ being \textit{maximal/minimal} anymore in \cite{VO1} and in the
Appendix of \cite{VO2}, and the changes of variables they employ are used
erroneously, as shown in the Introduction.

\bigskip

In the second step we prove that \textit{if} $\beta _{1}$ is \textit{%
maximal/minimal}, then $S_{11}\equiv \langle 1|\hat{U}\left( \infty ,-\infty
\right) |1\rangle $ does \textit{not} depend on $E\left( l_{j}\right) $, for
any $l_{j}=2,3,\ldots ,N$ and any $j=1,2,\ldots ,m$.

From Eq.$\left( \ref{fiveone}\right) $ it follows that: 
\begin{eqnarray}
\frac{\partial S_{11}}{\partial E\left( l_{j}\right) } &=&\sum_{m=1}^{\infty
}\left( -1\right) ^{m}\sum_{k_{1}=2}^{N}\sum_{k_{2}=2}^{N}\cdots
\sum_{k_{m}=2}^{N}\left\vert V_{1k_{1}}\right\vert ^{2}\left\vert
V_{1k_{2}}\right\vert ^{2}\cdots \left\vert V_{1k_{m}}\right\vert ^{2}\times
\nonumber \\
&&\times \int_{-\infty }^{\infty }d\tau _{1}\int_{-\infty }^{\tau _{1}}d\tau
_{2}\cdots \int_{-\infty }^{\tau _{2m-1}}d\tau _{2m}  \nonumber \\
&&\times i\left[ \delta \left( l_{j},k_{1}\right) \left( \tau _{1}-\tau
_{2}\right) +\delta \left( l_{j},k_{2}\right) \left( \tau _{3}-\tau
_{4}\right) +\cdots +\delta \left( l_{j},k_{m}\right) \left( \tau
_{2m-1}-\tau _{2m}\right) \right]  \nonumber \\
&&\times \exp \left[ i\frac{B\left( k_{1}\right) }{2}\left( \tau
_{1}^{2}-\tau _{2}^{2}\right) +i\frac{B\left( k_{2}\right) }{2}\left( \tau
_{3}^{2}-\tau _{4}^{2}\right) +\cdots +i\frac{B\left( k_{m}\right) }{2}%
\left( \tau _{2m-1}^{2}-\tau _{2m}^{2}\right) \right]  \nonumber \\
&&\times \exp \left[ iE\left( k_{1}\right) \left( \tau _{1}-\tau _{2}\right)
+iE\left( k_{2}\right) \left( \tau _{3}-\tau _{4}\right) +\cdots +iE\left(
k_{m}\right) \left( \tau _{2m-1}-\tau _{2m}\right) \right] ,  \label{eq3}
\end{eqnarray}%
where $\delta \left( l_{j},k_{p}\right) $ is the Kronecker delta.

Next, we introduce the well-known change of variables (see, e.g., textbook 
\cite{Fetter}, page 114): 
\begin{eqnarray}
\tau _{1} &=&x_{1}\in \left( -\infty ,\infty \right)  \nonumber \\
\tau _{j+1} &=&\tau _{j}-x_{j+1}\text{, \ \ with \ \ }x_{j+1}\in \left[
0,\infty \right) \text{, \ \ \ \ \ \ }j=1,2,\ldots ,2m-1\text{.}  \label{eq4}
\end{eqnarray}

From Eqs. $\left( \ref{eq3}\right) $ and $\left( \ref{eq4}\right) $ it
follows that 
\begin{eqnarray}
\frac{\partial S_{11}}{\partial E\left( l_{j}\right) } &=&\sum_{m=1}^{\infty
}\left( -1\right) ^{m}\sum_{k_{1}=2}^{N}\sum_{k_{2}=2}^{N}\cdots
\sum_{k_{m}=2}^{N}\left\vert V_{1k_{1}}\right\vert ^{2}\left\vert
V_{1k_{2}}\right\vert ^{2}\cdots \left\vert V_{1k_{m}}\right\vert ^{2}\times
\nonumber \\
&&\times \int_{-\infty }^{\infty }dx_{1}\int_{0}^{\infty
}dx_{2}\int_{0}^{\infty }dx_{3}\cdots \int_{0}^{\infty }dx_{2m}  \nonumber \\
&&\times i\left[ \delta \left( l_{j},k_{1}\right) x_{2}+\delta \left(
l_{j},k_{2}\right) x_{4}+\cdots +\delta \left( l_{j},k_{m}\right) x_{2m}%
\right]  \nonumber \\
&&\times F\left( x_{1},x_{2},x_{3},\ldots ,x_{2m}\right) ,  \label{eq5}
\end{eqnarray}%
where 
\begin{eqnarray}
F\left( x_{1},x_{2},x_{3},\ldots ,x_{2m}\right) &=&\exp \left[ -\frac{i}{2}%
G\left( x_{2},x_{4},\ldots ,x_{2m-2},x_{2m}\right) \right] \times  \nonumber
\\
&&\times \exp \left[ i\left\{ B\left( k_{1}\right) x_{2}+B\left(
k_{2}\right) x_{4}+\cdots +B\left( k_{m}\right) x_{2m}\right\} x_{1}\right] 
\nonumber \\
&&\times \exp \left[ -i\left\{ B\left( k_{2}\right) x_{4}+B\left(
k_{3}\right) x_{6}+\cdots +B\left( k_{m}\right) x_{2m}\right\} x_{3}\right] 
\nonumber \\
&&\times \exp \left[ -i\left\{ B\left( k_{3}\right) x_{6}+B\left(
k_{4}\right) x_{8}+\cdots +B\left( k_{m}\right) x_{2m}\right\} x_{5}\right] 
\nonumber \\
&&\vdots  \nonumber \\
&&\times \exp \left[ -i\left\{ B\left( k_{m}\right) x_{2m}\right\} x_{2m-1}%
\right] ,  \label{eq5-F}
\end{eqnarray}%
and 
\begin{eqnarray}
G\left( x_{2},x_{4},\ldots ,x_{2m-2},x_{2m}\right) &=&B\left( k_{1}\right)
\left( x_{2}^{2}\right) +B\left( k_{2}\right) \left(
2x_{4}x_{2}+x_{4}^{2}\right)  \nonumber \\
&&+B\left( k_{3}\right) \left( 2x_{6}x_{2}+2x_{6}x_{4}+x_{6}^{2}\right) 
\nonumber \\
&&\vdots  \nonumber \\
&&+B\left( k_{m}\right) \left( 2x_{2m}x_{2}+2x_{2m}x_{4}+\cdots
+2x_{2m}x_{2m-2}+x_{2m}^{2}\right)  \nonumber \\
&&-2\left[ E\left( k_{1}\right) x_{2}+E\left( k_{2}\right) x_{4}+\cdots
+E\left( k_{m}\right) x_{2m}\right] .  \label{eq5-G}
\end{eqnarray}

Upon integrating over $x_{1}$ in Eq.$\left( \ref{eq5}\right) $ one obtains: 
\begin{eqnarray}
\frac{\partial S_{11}}{\partial E\left( l_{j}\right) } &=&\sum_{m=1}^{\infty
}\left( -1\right) ^{m}\sum_{k_{1}=2}^{N}\sum_{k_{2}=2}^{N}\cdots
\sum_{k_{m}=2}^{N}\left\vert V_{1k_{1}}\right\vert ^{2}\left\vert
V_{1k_{2}}\right\vert ^{2}\cdots \left\vert V_{1k_{m}}\right\vert ^{2}\times
\nonumber \\
&&\times \int_{0}^{\infty }dx_{2}\int_{0}^{\infty }dx_{3}\cdots
\int_{0}^{\infty }dx_{2m}  \nonumber \\
&&\times i\left[ \delta \left( l_{j},k_{1}\right) x_{2}+\delta \left(
l_{j},k_{2}\right) x_{4}+\cdots +\delta \left( l_{j},k_{m}\right) x_{2m}%
\right]  \nonumber \\
&&\times \exp \left[ -\frac{i}{2}G\left( x_{2},x_{4},\ldots
,x_{2m-2},x_{2m}\right) \right]  \nonumber \\
&&\times 2\pi \delta \left( B\left( k_{1}\right) x_{2}+B\left( k_{2}\right)
x_{4}+\cdots +B\left( k_{m}\right) x_{2m}\right)  \nonumber \\
&&\times \exp \left[ -i\left\{ B\left( k_{2}\right) x_{4}+B\left(
k_{3}\right) x_{6}+\cdots +B\left( k_{m}\right) x_{2m}\right\} x_{3}\right] 
\nonumber \\
&&\times \exp \left[ -i\left\{ B\left( k_{3}\right) x_{6}+B\left(
k_{4}\right) x_{8}+\cdots +B\left( k_{m}\right) x_{2m}\right\} x_{5}\right] 
\nonumber \\
&&\vdots  \nonumber \\
&&\times \exp \left[ -i\left\{ B\left( k_{m}\right) x_{2m}\right\} x_{2m-1}%
\right] .  \label{eq6}
\end{eqnarray}

From Eq.$\left( \ref{eq2}\right) $ it follows that, for any $k_{j}$, $%
B\left( k_{j}\right) >0$ if $\beta _{1}$ is maximal, and $B\left(
k_{j}\right) <0$ if $\beta _{1}$ is minimal. Therefore, $\delta \left(
B\left( k_{1}\right) x_{2}+B\left( k_{2}\right) x_{4}+\cdots +B\left(
k_{m}\right) x_{2m}\right) =0$ unless $x_{2}=x_{4}=\cdots =x_{2m}=0$. The
presence of the term 
\begin{equation}
\left[ \delta \left( l_{j},k_{1}\right) x_{2}+\delta \left(
l_{j},k_{2}\right) x_{4}+\cdots +\delta \left( l_{j},k_{m}\right) x_{2m}%
\right]  \label{eq7}
\end{equation}
\noindent in the integrands of Eq.$\left( \ref{eq6}\right) $ ensures that
each of the integrals is zero.

This argument can be made rigorous by regularizing the behavior of integrals
at $\pm \infty $, as follows 
\begin{eqnarray}
&&I_{q}\equiv \int_{-\infty }^{\infty }dx_{1}\int_{0}^{\infty
}dx_{2}\int_{0}^{\infty }dx_{3}\cdots \int_{0}^{\infty }dx_{2m}\text{ }%
F\left( x_{1},\ldots ,x_{2m}\right) x_{2q}  \nonumber \\
&&=\lim_{\eta \rightarrow 0_{+}}\left\{ \text{ }\int_{0}^{\infty
}dx_{2}\int_{0}^{\infty }dx_{4}\cdots \int_{0}^{\infty
}dx_{2m}\int_{0}^{\infty }dx_{1}\int_{0}^{\infty }dx_{3}\cdots
\int_{0}^{\infty }dx_{2m-1}\right. \times  \nonumber \\
&&\times x_{2q}F\left( x_{1},\ldots ,x_{2m}\right) e^{-\eta \left(
x_{1}+x_{3}+\cdots +x_{2m-1}\right) }  \nonumber \\
&&+\int_{0}^{\infty }dx_{2}\int_{0}^{\infty }dx_{4}\cdots \int_{0}^{\infty
}dx_{2m}\int_{-\infty }^{0}dx_{1}\int_{0}^{\infty }dx_{3}\cdots
\int_{0}^{\infty }dx_{2m-1}\times  \nonumber \\
&&\times \left. x_{2q}F\left( x_{1},\ldots ,x_{2m}\right) e^{-\eta \left(
-x_{1}+x_{3}+\cdots +x_{2m-1}\right) }\text{ }\right\} ,  \label{eq8}
\end{eqnarray}
\noindent for any $q=1,2,\ldots ,m$. Upon integrating over the $x_{j}$ with
odd $j$ in Eq.$\left( \ref{eq8}\right) $, one obtains 
\begin{equation}
I_{q}=\lim_{\eta \rightarrow 0_{+}}\int_{0}^{\infty }dx_{2}\int_{0}^{\infty
}dx_{4}\cdots \int_{0}^{\infty }dx_{2m}\text{ }W\left( x_{2},x_{4},\ldots
,x_{2m};\eta \right) ,  \label{eq9}
\end{equation}
\noindent where 
\begin{eqnarray}
W\left( x_{2},x_{4},\ldots ,x_{2m};\eta \right) &=&\exp \left[ -\frac{i}{2}%
G\left( x_{2},x_{4},\ldots ,x_{2m}\right) \right] \times  \nonumber \\
&&\times \frac{2\eta x_{2q}}{\left[ B\left( k_{1}\right) x_{2}+B\left(
k_{2}\right) x_{4}+\cdots +B\left( k_{m}\right) x_{2m}\right] ^{2}+\eta ^{2}}
\nonumber \\
&&\times \frac{1}{i\left[ B\left( k_{2}\right) x_{4}+B\left( k_{3}\right)
x_{6}+\cdots +B\left( k_{m}\right) x_{2m}\right] +\eta }  \nonumber \\
&&\vdots  \nonumber \\
&&\times \frac{1}{i\left[ B\left( k_{m}\right) x_{2m}\right] +\eta }.
\label{eq10}
\end{eqnarray}

Since either $B\left( k_{j}\right) >0$ or $B\left( k_{j}\right) <0$ for any $%
k_{j}$, from Eqs. $\left( \ref{eq9}\right) $ and $\left( \ref{eq10}\right) $
it follows that 
\begin{eqnarray}
\left\vert I_{q}\right\vert &\leq &\lim_{\eta \rightarrow
0_{+}}\int_{0}^{\infty }dx_{2}\int_{0}^{\infty }dx_{4}\cdots
\int_{0}^{\infty }dx_{2m}\text{ }\left\vert W\left( x_{2},x_{4},\ldots
,x_{2m};\eta \right) \right\vert  \nonumber \\
&=&\lim_{\eta \rightarrow 0_{+}}\int_{0}^{\eta /\left\vert B\left(
k_{1}\right) \right\vert }dx_{2}\int_{0}^{\eta /\left\vert B\left(
k_{2}\right) \right\vert }dx_{4}\cdots \int_{0}^{\eta /\left\vert B\left(
k_{m}\right) \right\vert }dx_{2m}\times  \nonumber \\
&&\times \frac{2\eta x_{2q}}{\left[ B\left( k_{1}\right) x_{2}+B\left(
k_{2}\right) x_{4}+\cdots +B\left( k_{m}\right) x_{2m}\right] ^{2}+\eta ^{2}}
\nonumber \\
&&\times \frac{1}{\sqrt{\left[ B\left( k_{2}\right) x_{4}+B\left(
k_{3}\right) x_{6}+\cdots +B\left( k_{m}\right) x_{2m}\right] ^{2}+\eta ^{2}}%
}\cdots \frac{1}{\sqrt{\left[ B\left( k_{m}\right) x_{2m}\right] ^{2}+\eta
^{2}}}.  \label{eq11}
\end{eqnarray}

For any continuous function $f$ defined on a compact interval $\left[ a,b%
\right] $ the relation $\int_{a}^{b}f\left( x\right) dx=f\left( \xi \right)
\left( b-a\right) $ holds, for some $\xi \in \left[ a,b\right] $. Hence, Eq.$%
\left( \ref{eq11}\right) $ reduces to 
\begin{eqnarray}
\left\vert I_{q}\right\vert &\leq &\frac{1}{\left\vert B\left( k_{1}\right)
B\left( k_{2}\right) \cdots B\left( k_{m}\right) \right\vert }\lim_{\eta
\rightarrow 0_{+}}2\eta \frac{\xi _{q}}{\left\vert B\left( k_{q}\right)
\right\vert }\frac{1}{\left( \xi _{1}+\xi _{2}+\cdots +\xi _{m}\right) ^{2}+1%
}\times  \nonumber \\
&&\times \frac{1}{\sqrt{\left( \xi _{2}+\xi _{3}+\cdots +\xi _{m}\right)
^{2}+1}}\cdots \frac{1}{\sqrt{\xi _{m}^{2}+1}}=0,  \label{eq12}
\end{eqnarray}%
\noindent where $\xi _{j}\in \left[ 0,1\right] $, for any $j=1,2,\ldots ,m$.
Eq.$\left( \ref{eq12}\right) $ shows that $I_{q}=0$ for any $q=1,2,\ldots ,m$%
, and consequently $\frac{\partial S_{11}}{\partial E\left( l_{j}\right) }=0$
for any $E\left( l_{j}\right) $.

The fact that $\beta _{1}$ is \textit{maximal/minimal} has played a key role
in the arguments above. These arguments fail to hold if $\beta _{1}$ is not
an extremum, since then the argument of $\delta \left( B\left( k_{1}\right)
x_{2}+B\left( k_{2}\right) x_{4}+\cdots +B\left( k_{m}\right) x_{2m}\right) $
would have other zeros, besides the obvious $x_{2}=x_{4}=\cdots =x_{2m}=0$.

\bigskip

In the last step, we reveal the connection between the BEF for a general $N$%
-level LZ-system and the exactly solvable bow-tie model \cite{bow-tie}, in
which all levels interact with only one special level (SL). Indeed, since $%
S_{11}$ does not depend on $E\left( l_{j}\right) $ \textit{if} $\beta _{1}$
is \textit{maximal/minimal}\textbf{,} we can safely set $E\left(
l_{j}\right) =0$ for any $l_{j}=2,3,\ldots ,N$. The form for $S_{11}$ (Eq.$%
\left( \ref{fiveone}\right) $) with all $E\left( l_{j}\right) =0$ is exactly
what one obtains for a bow-tie model if the SL has slope $\beta _{1}$, and a
perturbation expansion is being developed for it.

Since the Brundobler-Elser conjecture for the bow-tie model was proven
analytically by Ostrovsky and Nakamura in \cite{bow-tie}, we only cite their
statement from page 6947 of \cite{bow-tie}: \textquotedblleft \textit{This
hypothesis is confirmed within the present model.}\textquotedblright , and
refer to their work for further details.

This ends the proof of the Brundobler-Elser formula for a general $N$-level
LZ-system. Before drawing the conclusions, it is worthwhile to mention that,{%
\ }preceding \cite{VO1}, {an alternative proof of BEF, based on analytic
continuation in imaginary time, was proposed by Shytov in \cite{Shytov}.}

\bigskip


\section{Conclusions}

In summary, we revealed the deficiencies of the BEF proof proposed by Volkov
and Ostrovsky in \cite{VO1} and Appendix of \cite{VO2}. We emphasized that
using the important property of the slope $\beta _{1}$ being an extremum 
\textit{only} to arrive at the sub-series $(5.1)$ in their proof is \textit{%
not enough}, regardless of the changes of variables they attempted, and this
property still has to play a crucial role in evaluating $(5.1)$.

We have constructed a proof of BEF that corrects for this shortcoming.
Indeed, in our proof, the fact that the slope $\beta _{1}$ is
maximal/minimal had to be used for three different purposes: 1) to arrive at
Eq.$(5.1)$ as was shown in \cite{VO1} ; 2) to prove that $S_{11}$ does not
depend on $E\left( l_{j}\right) $; 3) to employ the fact that the
Brundobler-Elser conjecture for maximum/minimum slope was proved for the
particular case of the bow-tie model \cite{bow-tie}.

\bigskip


\textit{Note added.-}After this work was submitted to publication we
received a Reply by Volkov and Ostrovsky \cite{VO-Reply} in which these
authors now verify, by direct calculation, the important point on which they
missed before in \cite{VO1} and Appendix of \cite{VO2}, and which we have
advocated throughout our Comment, namely that in order for the BEF to hold
the important property of $\beta _{1}$ being an \textit{extreme slope} (%
\textit{ES}) must be used manifestly in the evaluation of \textit{series} $%
(5.1)$.

In their Reply V-O abandon the old changes of variables and use this time
the same well-known change of variables $\left( \ref{eq4}\right) $ (see,
e.g., textbook \cite{Fetter}, page 114) as we have employed. We note that
this change of variables, while previously used by V-O in a different
context (e.g., in the proof of the no-go theorem \cite{VO2}), was never
employed in their attempts of calculating \textit{series }$(5.1)$ in \cite%
{VO1} and Appendix of \cite{VO2}. The point at issue in our Comment is,
however, not about what changes of variables should be or not be used in the
proof of BEF. Mathematical inconsistencies aside, it was definitely not the
V-O's choice of integration variables (in \cite{VO1} and Appendix of \cite%
{VO2}) that prompted our Comment, but their \textit{total neglect} of the
very important \textit{ES}-property of $\beta _{1}$ in evaluating \textit{%
series }$(5.1)$.

Though only the evaluation of 4-th order terms is presented in \cite%
{VO-Reply}, our claim that one can set $\varepsilon _{j}=0$ \textit{only if} 
$\beta _{1}$ is an \textit{ES} is reconfirmed, together with the fact that
this \textit{ES}-property must \textit{still} be used even after setting $%
\varepsilon _{j}=0$ in order to prove the Brundobler-Elser conjecture.

\bigskip


\section{Acknowledgments}

We thank V. L. Pokrovsky for encouragement. This work was supported by DOE
grants DE-FG03-96ER45598 and DE-FG03-02ER45958, by NSF under the grants
DMR-0115947 and DMR-0321572, and by the Welch Foundation.


\begin{thebibliography}{9}
\bibitem{VO1} Volkov M V and Ostrovsky V N 2004 \textit{J. Phys. B: At. Mol.
Opt. Phys.} \textbf{37} 4069-84

\bibitem{VO2} Volkov M V and Ostrovsky V N 2005 \textit{J. Phys. B: At. Mol.
Opt. Phys.} \textbf{38} 907-15

\bibitem{BE} Brundobler S and Elser V 1993 \textit{J. Phys. A: Math. Gen.} 
\textbf{26} 1211-27

\bibitem{bow-tie} Ostrovsky V N and Nakamura H 1997 \textit{J. Phys. A:
Math. Gen.} \textbf{30} 6939-50

\bibitem{KayaF} Kayanuma Y and Fukuchi S 1985 \textit{J. Phys. B: At. Mol.
Phys.} \textbf{18} 4089-4093

\bibitem{Shytov} Shytov A V 2004 \textit{Phys. Rev. A} \textbf{70} 052708

\bibitem{Fetter} Fetter A L and Walecka J D 1971 \textit{Quantum Theory of
Many-Particle Systems} (McGraw-Hill, New York)

\bibitem{VO-Reply} Volkov M V and Ostrovsky V N 2005 \textit{J. Phys. B: At.
Mol. Opt. Phys.}, this volume
\end{thebibliography}
\end{document}